\documentclass[aps,arl,groupedaddress,twocolumn]{revtex4}

\usepackage{graphicx}
\usepackage{amsmath}
\usepackage{amsfonts}
\usepackage{amssymb}

\usepackage{subfigure}

\usepackage{color}

\begin{document}


\title{
Competition between loss channels in quantum-dot cavity systems: \\ unconventional consequences
}

\author{A. Vagov$^{1}$}
\email[]{alexei.vagov@uni-bayreuth.de}
\author{M. Gl\"assl$^{1}$}
\author{M. D. Croitoru$^{1,2}$}
\author{V. M. Axt$^{1}$}
\author{T. Kuhn$^{3}$}

\affiliation{$^{1}$Institut f\"{u}r Theoretische Physik III,
Universit\"{a}t Bayreuth, 95440 Bayreuth, Germany}
\affiliation{$^{2}$Departement Fysica, Universiteit Antwerpen,
Groenenborgerlaan 171, B-2020 Antwerpen, Belgium}
\affiliation{$^{3}$Institut f\"ur Festk\"orpertheorie, Westf\"alische
Wilhelms-Universit\"at M\"unster, Germany}

\date{\today}

\begin{abstract}
We demonstrate that in quantum-dot cavity systems, the interplay between acoustic phonons 
and photon losses introduces novel features and characteristic dependencies in the system
dynamics. In particular, the combined action of both loss mechanisms strongly affects 
the transition from the weak to the strong coupling regime as well as the shape of 
Mollow-type spectra in untypical ways. For weak coupling, where the spectra
degenerate to a single line, we predict that their widths decrease with rising temperature.
\end{abstract}

\maketitle


High quality quantum dot (QD) cavity systems allow the realization
of various technologically important devices such as sources
for single \cite{press:07} or entangled photons \cite{dousse:10},
that are relevant for applications in quantum information
processing \cite{bouwmeester:00} as well as for tests of fundamental
aspects of quantum mechanics \cite{haroche:06}.  
As state-of-the-art cavities \cite{boeckler:08} have now reached a quality, where the unavoidable 
phonon related loss channels can compete with cavity losses, 
studies of the phonon-influence on these systems has become a major focus
of topical research. It has been shown that the QD acoustic
phonon coupling is responsible for an unexpectedly strong QD-cavity coupling
for cavities that are detuned from the QD resonance \cite{ates:09, kaer:10},
which manifests itself in a spectral broadening of the Mollow sidebands through 
off-resonant cavity emission \cite{ulrich:11, roy:11}. 
Further, it was demonstrated that phonons strongly influence 
the photon statistics \cite{kaer:13a,harsij:12,glaessl:12b} and  
lead to an enhanced coherent scattering as well as to
an off-resonant sideband narrowing \cite{nazir:13, ulrich:11}.
Recently, we have performed a numerically complete path-integral 
analysis dealing with the limiting case
of negligible cavity losses and demonstrated that in this case, due to the phonon coupling, 
an increase of the light-matter coupling can counter-intuitively
reduce the visibility of Jaynes-Cummings revivals and lead to a broadening
and lowering of spectral sidebands \cite{glaessl:12b}.  

In this Letter, we demonstrate that the competition of the phonon-induced decoherence
with cavity losses brings in qualitatively new aspects to the system dynamics 
compared to the case when one of the two mechanisms is absent. 
In particular, the interplay between phonon-induced pure dephasing and phonon-induced renormalizations
of the light-matter coupling with cavity and radiative losses strongly 
and non-trivially affects the  transition from the weak to the strong coupling regime as well as
Mollow-type spectra, which is manifested in characteristic dependencies on temperature and  
the light-matter coupling strength. 
Our analysis combines a microscopic calculation of phonon induced pure dephasing effects
with a phenomenological treatment of other loss channels providing an easy link to
the standard analysis of the cross-over between the weak and strong coupling regimes
which in the literature is usually discussed without accounting for pure dephasing.

In many typical QD-cavity experiments, the mean photon number is
below one. Here, we shall concentrate on this case, which implies that we have
to account for the dynamics only in the subspace spanned by the three states: 
$|0\rangle = |G, n=0\rangle$, 
$|1\rangle = |X, n=0\rangle$ 
and 
$|2\rangle = |G, n=1\rangle$, 
where $G$ denotes the empty dot without an exciton, while $X$ marks the exciton
state. The integer label $n$ refers to the number of photons that are excited 
in the cavity mode. Other dot states are not considered as they are assumed to
be coupled off-resonantly. Using the Jaynes-Cummings model to describe the coupling
between the states $|1\rangle$ and $|2\rangle$, closed equations are obtained for 
the evolution of the density matrix operator $\hat\rho_{\ell j}=|\ell\rangle\langle j|$
and the corresponding expectation values $\rho_{\ell j}=\langle\hat\rho_{\ell j}\rangle$.
The simplest way  to account for cavity losses and the dot-environment coupling 
by radiative decay is to introduce phenomenological rates into the equations of
$\rho_{\ell j}$ following the Lindblad formalism as described in standard text
books \cite{breuer:02}. The goal of the present paper is to provide a clear analysis
of the competition of these decay mechanisms with phonon-induced pure dephasing  
and to characterize the resulting interplay of all loss channels in a comprehensive
way. 

For a cavity in resonance with the dot transition, a phenomenological model that
accounts for all of the  above loss channels can be formulated as follows
\footnote{The dynamics of the density matrix
elements involving the ground state $|0\rangle$ do not couple back to the
above variables and thus need not to be considered here.}:

\begin{align}
 \frac{d}{dt} \left( \begin{array}{c}
         \rho_{11} \\
         \rho_{22} \\
          V 
        \end{array}
\right) \!=\! 
\underbrace{
\left( \begin{array}{ccc}
                  -r & 0 & i g  \\
                  0 & - \kappa & - i g \\
                  2i\tilde{g} & -2i\tilde{g} & -(\kappa+r)/2 - \gamma 
                 \end{array} 
           \right)
}_{=:M}
\!\! \left( \begin{array}{c}
         \rho_{11} \\
         \rho_{22} \\
          V 
        \end{array}
\right),
\label{rate}
\end{align} 
where $V = \rho_{12} - \rho_{21}$.
$\kappa$ accounts for cavity losses and can be
expressed in terms of the quality factor $Q$ and the dot transition frequency
$\omega_{0}$ as $\kappa=\omega_{0}/Q$. 
$r$ is the radiative decay rate, while 
$\gamma$ stands for the phonon-induced pure dephasing.
$g$ denotes the bare QD-cavity coupling
strength, which in the loss-free case is related to the Rabi frequency
$\Omega$ by $\Omega=2g$. Finally, 
$\tilde{g}$ represent a renormalized light-matter coupling
which models the phonon-induced renormalization of the Rabi frequency 
known from microscopic treatments \cite{kaer:10,glaessl:12b}.

The pure dephasing rate $\gamma$ is introduced according to  the Lindblad formalism \cite{nazir:13}.
Note, that $\gamma$, unlike $\kappa$ and $r$, does not enter the equations for
$\rho_{11}$ and $\rho_{22}$ as pure dephasing alone cannot change
occupations \cite{mahan:90,mukamel:95,krummheuer:02}.
The Lindblad formalism, however, does not account for frequency renormalizations.
As a phenomenological way to describe these renormalizations,
we have introduced $\tilde{g}$ in our model \footnote{In accordance with the microscopic
treatment, the phonon-induced modifications enter only in the equation for
the off-diagonal density matrix elements.}.

An overview of the behavior of this dissipative system is most easily
obtained by looking at the eigenvalues of the matrix $M$ in Eq.~\eqref{rate}.
While analytic expressions for these eigenvalues are available, in the general
case, when all rates in Eq.~\eqref{rate} are non-zero, they are quite lengthy
and not instructive. This is different in two limiting cases. 
First, when pure dephasing is negligible, i.e., in the limit  $\gamma=0$ and $\tilde{g}=g$, 
we recover the well known result \cite{andreani:99, rudin:99}:
\begin{align}
  \label{g0}
  \lambda_{0} = -\frac{\kappa+r}{2},\qquad
  \lambda_{\pm} = \lambda_{0} \pm \frac{1}{2} \sqrt{(\kappa-r)^{2}-16g^{2}}.
\end{align}
The $\lambda_{0}$ eigenmode decays with the average rate of cavity losses
and radiative decay and never shows oscillations. The $\lambda_{\pm}$ modes
become complex for $4g>|\kappa-r|$, i.e., for large enough $g$, these modes
oscillate and exhibit the same $g$-independent damping as the $\lambda_{0}$
mode. Thus, the $\lambda_{\pm}$ modes correspond to the side-peaks in
Mollow-type spectra, while $\lambda_{0}$ contributes to the unshifted
central peak. Eq.~\eqref{g0} predicts that the bifurcation point that
separates the weak from the strong coupling regime can even for finite
$\kappa$ be shifted to arbitrarily small values of $g$ when $\kappa$ and $r$
are similar. However, in typical QD-cavity systems, $\kappa$ dominates by far. 

The second simple limiting case is obtained by setting $\kappa=r=0$ and
keeping $\gamma$ finite. This situation is usually not discussed, as typically
$\kappa$ cannot be neglected and phonon-induced pure dephasing is regarded
to be marginal, especially at low temperatures. Nevertheless, for later
comparisons, it is instructive to shortly look at this case, where the
eigenvalues are given by
\begin{align}
  \label{kr0}
  \lambda_{0} = 0,\qquad
  \lambda_{\pm} = -\frac{\gamma}{2} \pm \frac{1}{2} \sqrt{\gamma^{2}-16g\tilde{g}},
\end{align}
revealing that pure dephasing alone does not damp the $\lambda_{0}$ mode but 
only the $\lambda_{\pm}$ modes. For any finite and constant
$\gamma$, we obtain a well defined bifurcation point at $16g\tilde{g}=\gamma^{2}$ and for larger 
$g\tilde{g}$, the damping of the oscillating modes becomes independent of the light-matter coupling 
and is given  by $\gamma/2$.

\begin{figure*}[t]%
\includegraphics[width=17.0cm]{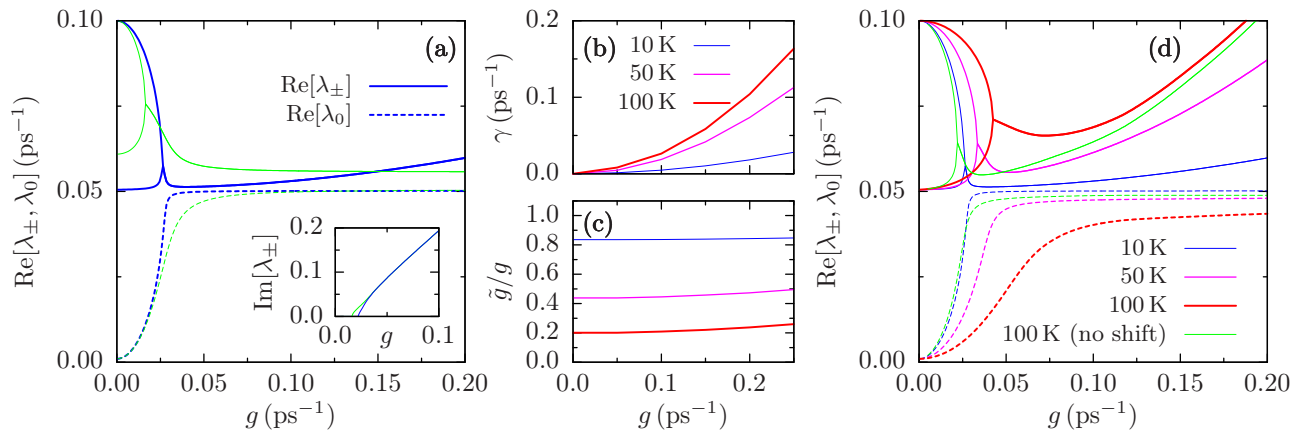}
\caption{(Color online) 
Panel (a):  real parts of the eigenvalues of the matrix $M$ in Eq.~\eqref{rate} as a function
of the bare light-matter coupling $g$ for $\kappa=0.1\,\rm{ps}^{-1}$ and $r=0.001\,\rm{ps}^{-1}$
assuming a constant pure dephasing rate of $\gamma=0.01\,\rm{ps}^{-1}$ (green thin lines)
and using the exact dephasing rate $\gamma(g)$ and accounting for the renormalization
$\tilde{g}(g)$ as obtained for a spherical GaAs dot with radius 3~nm at $T=10$~K (blue thick lines).
Inset: corresponding imaginary parts of $\lambda_{+}$ ($\rm{Im}[\lambda_0]=0$, 
$\rm{Im}[\lambda_-] = -\rm{Im}[\lambda_+]$). 
Panels (b) and (c):  pure dephasing rate $\gamma(g)$ and renormalization of the light-matter coupling
$\tilde g (g)/g$ as obtained from exact path integral calculations at three temperatures. 
Panel (d): real parts of the eigenvalues of the matrix $M$ in Eq.~\eqref{rate} at $T=10$, 50 and 100~K calculated
using the exact damping rate $\gamma(g)$ as well as accounting for the the renormalization $\tilde g(g)$
for $\kappa=0.1\,\rm{ps}^{-1}$ and $r=0.001\,\rm{ps}^{-1}$. 
The green thin line gives results as obtained for $T=100$~K accounting for $\gamma(g)$ but
assuming $\tilde{g} = g$. 
} 
\label{fig1}
\end{figure*}

Interestingly, when $\kappa$, $r$ and $\gamma$ are all non-zero, the situation
changes qualitatively compared with the limiting cases of Eqs.~\eqref{g0} and 
\eqref{kr0}, even when we neglect the frequency renormalization and put $\tilde{g}=g$. 
The green thin lines in Fig.~\ref{fig1}(a) display the eigenvalues 
of $M$ calculated for $\kappa=0.1\,\rm{ps}^{-1}$, $r=0.001\,\rm{ps}^{-1}$, and $\gamma=0.01\,\rm{ps}^{-1}$
as functions of $g=\tilde{g}$. In contrast to Eqs.~\eqref{g0} and \eqref{kr0}, the real
parts that correspond to the damping of the modes show a noticeable dependence
on $g$ in the strong coupling regime, even though all rates are taken independent
on $g$. In particular, $Re[\lambda_{\pm}]$ decreases with rising $g$ for $g$
above the bifurcation point. The competition between the different decay channels
also strongly affects the position of the bifurcation point. Here, it is essentially
only the competition between $\kappa$ and $\gamma$, as $r$ is too small to have a
noticeable influence. For the chosen parameters, the bifurcation point is found at
$g_{b}\simeq 0.017 \,\rm{ps}^{-1}$, which should be compared with the bifurcation point
defined by Eq.~\eqref{g0} that is given by
$\overline{g}_{b}=|\kappa-r|/4\simeq 0.025\,\rm{ps}^{-1}$, i.e., adding a constant  pure
dephasing rate to the cavity loss shifts the bifurcation point to lower values, similar
to the effect of a finite value of $r$ in Eq.~\eqref{g0}.

So far, our analysis assumed rates that are independent of $g$.
>From a microscopic point of view, these assumptions need to be
revisited. First of all, pure dephasing is a non-Markovian effect,
as can be seen from the fact that the phonon-induced 
polarization decay  after a short pulse excitation is only partial and 
non-exponential \cite{krummheuer:02}. 
In this Letter, however, we study the
case of a cavity  with at most a single photon, which is formally equivalent
to a two-level QD without cavity, subject to laser driving with a constant amplitude.
Here, the non-Markovian nature of the coupling manifests itself in
non-vanishing  off-diagonal elements of $\rho$ at long times \cite{glaessl:11b}.
However, for resonant driving, the non-vanishing part of $\rho_{12}$ equals
$\text{Re}[\rho_{12}]$, which in this case is dynamically decoupled 
from the three dynamical variables $\rho_{11}$, $\rho_{22}$ and $V$
appearing in Eq.~(\ref{rate}).
The phonon-induced pure dephasing has
recently been studied for QD-cavity systems \cite{glaessl:12b} within a
numerically complete path integral approach \cite{vagov:11} in the limit of
vanishing cavity and radiative losses. 
Considering a situation with at most one photon, 
the dynamics is restricted to a two-level system
and it turns out that for a cavity in resonance with the dot transition,
the path integral results  for the three dynamical variables $\rho_{11}$, $\rho_{22}$ and $V$
agree perfectly with those obtained from Eqs.~(\ref{rate}), provided
$\gamma$ and $\tilde{g}$ are suitably chosen.

In order to extract $\gamma$ and $\tilde{g}$ from our path integral
calculations, we consider the solution for 
a system initially prepared in the state $|2\rangle$, 
where the exciton occupation resulting from the simulation 
can be perfectly described by the formula 
$\rho_{X}=\rho_{11}=\frac{1}{2}\big[1-e^{-\Gamma t}\cos(\omega t)\big]$.
The fitting parameters $\Gamma$ and $\omega$ depend on $g$, the temperature $T$,
and the dot parameters that enter the carrier-phonon coupling. Further, it was
found from the path integral analysis, 
that for $\kappa=r=0$, the system exhibits
oscillations for all $g$. This finding of oscillatory solutions in the
microscopic theory for arbitrary small $g$ implies that the $g$ dependent
renormalizations of $\Gamma$ and $\omega$ prevent the system from having a
crossover to the weak coupling limit, i.e., unlike the theory with a $g$ independent
pure dephasing rate [cf.~\eqref{kr0}], there is no bifurcation point in a model
where phonon-induced pure dephasing is the only loss channel. From Eq.~\eqref{kr0}
we learn that for $\kappa=r=0$, the rate equation Eq.~\eqref{rate} yields the same
damped oscillation as the microscopic theory, provided we identify $\gamma=2\Gamma(g)$
and replace in Eq.~\eqref{rate} the bare light-matter coupling $g$ by
$\tilde{g} =[\Gamma^{2}(g)+\omega^{2}(g)]/(4g)$. 

As the light-matter coupling also enters the radiative recombination rate, 
$r$ should exhibit a corresponding $g$ dependence, that we will, however, neglect
due to the overall smallness of $r$. Cavity losses on the other hand are mainly
properties of the cavity and hence, no significant influence of $g$ is expected.
Therefore, we shall in the following use the model in Eq.~\eqref{rate} with
$\gamma(g)$ and $\tilde{g}(g)$ determined by our microscopic path-integral approach
and keep $\kappa=0.1\,\rm{ps}^{-1}$ and $r=0.001\,\rm{ps}^{-1}$ fixed.

$\gamma(g)$ and $\tilde{g}(g)$ are plotted in Figs.~\ref{fig1}(b,c) for a spherical 
GaAs QD with radius 3 nm and for three temperatures \footnote{The carrier-phonon
coupling constants are taken from Ref.~\onlinecite{krummheuer:02}}.  
Most important for our present discussion is that $\gamma(g)$ approaches
zero for $g\to0$ and rises with rising $T$, while $\tilde{g}(g)$ exhibits pronounced
and strongly $T$-dependent renormalizations. Accounting for these
renormalizations in Eq.~\eqref{rate} has significant consequences on the $g$
dependence of the eigenvalues $\lambda_{\pm,0}$, as demonstrated by the blue
thick lines in Fig.~\ref{fig1}(a) calculated for $T=10$~K. Most striking is that
the bifurcation point is now slightly above the value expected from Eq.~\eqref{g0}
(i.e. vanishing phonon influence) and not below as it is expected for a constant
$\gamma$ (cf. the green thin lines). The same trend is seen when looking at the
thick lines in Fig.~\ref{fig1}(d) that display the eigenvalues $\lambda_{\pm,0}$
for different temperatures accounting for the $g$ dependencies of $\gamma$ and
$\tilde{g}$. Obviously, the bifurcation point shifts to higher $g$ values with
rising temperatures. It is interesting to note, that when only the $g$ dependence
of $\gamma$ is taken into account, the opposite trend is found [cf. the thin lines
in Fig.~\ref{fig1}(d)]. The latter result is due to the fact that with rising $T$, the
phonon-induced damping $\gamma$ increases [cf. Fig.~\ref{fig1}(b)] and becomes comparable
to $\kappa$, resulting in a partial compensation as discussed before for a constant
$\gamma$ in Fig.~\ref{fig1}(a) and similar to the competition between $\kappa$ and $r$
in Eq.~\eqref{g0}. However, this partial compensation arising from the competition
between $\kappa$ and $\gamma(g)$ is overcompensated by the renomarlization of $g$,
the stronger, the higher the temperature is, and this overcompensation leads to a shift
of the bifurcation point towards larger light-matter couplings.

Another striking feature of the full theory with all g-dependent phonon-induced
renormalizations is that above the bifurcation point, we now find that the damping
of the oscillatory modes $\lambda_{\pm}$ increases with rising $g$, which is opposite
to the trend seen in calculations with a $g$-independent $\gamma$  [green thin lines
in Fig.~\ref{fig1}(a)]. The $g$ range displayed in Figs.~\ref{fig1}(a,d)
represents roughly the range of values for $g$ that has been realized in typical 
state-of-the-art cavities \cite{reithmaier:04, hennessy:07}. We note in passing,
that when $g$ is further increased, $\rm{Re}[\lambda_{\pm}]$ eventually decreases again,
mainly due to the non-monotonic $g$-dependence of $\gamma$ arising from the resonant
nature of the carrier-phonon coupling \cite{machnikowski:04,vagov:07,glaessl:12b}.

\begin{figure}[t]%
\includegraphics[width=7.6cm]{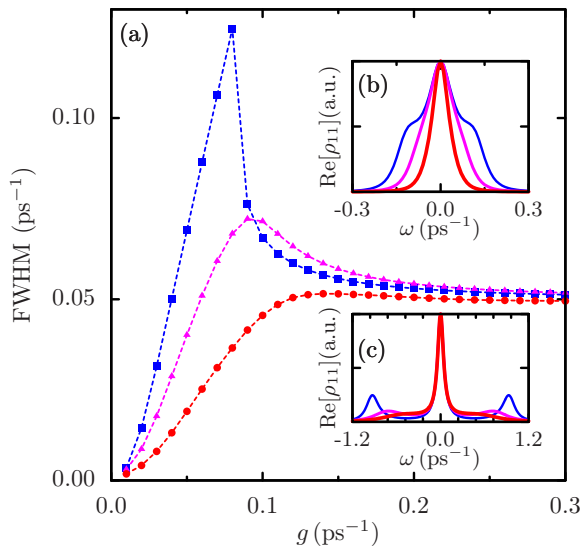}
\caption{(Color online) 
Panel (a): Full width at half maximum  (FWHM) of the central peak in the Mollow-type spectra 
${\rm Re}[\rho_{X}(\omega)]$ as a function of the light-matter coupling $g$ calculated at
$T=10$~K (blue dashed line with square symbols), $T=50$~K (purple dashed line with triangles)
and $T=100$~K (red dashed line with circles). Panels (b) and (c): spectra at $T=10$ (thin blue line),
50 (purple line) and 100~K (thick red line) calculated for (b) $g=0.075\,\rm{ps}^{-1}$
and (c) $g=0.15\,\rm{ps}^{-1}$, respectively. In all pannels, $\kappa=0.1\,\rm{ps}^{-1}$ and $r=0.001\,\rm{ps}^{-1}$.}  
\label{fig2}
\end{figure}

Finally, let us turn to the $\lambda_0$ mode. Fig.~\ref{fig1}(d) clearly shows
that the damping of the $\lambda_{0}$ mode increases with rising $g$, but that
it decreases with rising temperature. We stress that this counterintuitive temperature dependence
is introduced only by the competition between cavity losses and
phonon-induced pure dephasing: cavity losses alone would not lead to a $T$ dependence
in our model, as $T$ enters only in the phonon-induced parts. Pure dephasing
alone would also not lead to a $T$ dependence, as Eq.~\eqref{kr0} predicts
no damping of the $\lambda_{0}$ mode at all. 

The temperature dependence resulting from the competition between $\kappa$ and
$\gamma(g)$ also affects observable quantities like the Full-Width-at-Half-Maximum
(FWHM) of the peak centered at $\omega=0$  in the spectra $Re[\rho_{X}(\omega)]$ 
that are extracted in Fig.~\ref{fig2}(a) as a function of $g$ for different $T$. 
It should be noted that also the $\lambda_{\pm}$ modes contribute to the width
of that peak as long as the corresponding frequencies are not well separated from
$\omega=0$ [cf. Figs.~\ref{fig2}(b,c)]. For $T$=100~K, the heights of the side
peaks is generally low and thus the FWHM in Fig.~\ref{fig2}(a) reflects only the
monotonically rising contribution of the $\lambda_{0}$ mode. For lower temperatures, 
a maximum appears because at low $g$, the $\lambda_{\pm}$ contributions strongly
enhance the FWHM, while their influence decreases with rising $g$, when the side 
peaks move to higher $\omega$ and eventually become clearly separated. In the
weak coupling limit, the $\lambda_{\pm}$ contributions are also centered at
$\omega=0$ and their damping typically rises with rising temperature. This trend
is, however, overcompensated by the decreasing damping of the $\lambda_{0}$ mode,
resulting altogether in a decrease of the FWHM with rising $T$. Due to the competition
between the $\lambda_{0}$ and the $\lambda_{\pm}$ contributions, the $T$ dependence
of the FWHM becomes non-monotonic at intermediate $g$, while at high $g$, where the
$\lambda_{\pm}$ modes are not contributing any more, the low $g$ trend, reflecting the 
dependence of the $\lambda_{0}$ mode, is restored. 

In conclusion we have shown that the interplay between cavity losses and the $g$
dependent pure dephasing as well as the phonon-induced renormalization of the bare
light-matter coupling strength $g$ strongly affect the transition from the weak to
the strong coupling regime, e.g., reflected in $T$ dependent shifts of the bifurcation
point. 
In the weak coupling limit, we predict that the width of the peak at $\omega=0$
decreases for higher temperatures. Our analysis reveals that the temperature
dependence of this line results from the combined action of cavity losses and
phonon-induced pure dephasing; both mechanisms alone would lead to a temperature
independent width.  

We acknowledge fruitful discussions with A. Nazir which helped to
more clearly formulate the relation between our phenomenological approach and
the microscopic theory. M.D.C. further acknowledges Alexander von Humboldt
and BELSPO grants for support.


\begin{thebibliography}{25}
\expandafter\ifx\csname natexlab\endcsname\relax\def\natexlab#1{#1}\fi
\expandafter\ifx\csname bibnamefont\endcsname\relax
  \def\bibnamefont#1{#1}\fi
\expandafter\ifx\csname bibfnamefont\endcsname\relax
  \def\bibfnamefont#1{#1}\fi
\expandafter\ifx\csname citenamefont\endcsname\relax
  \def\citenamefont#1{#1}\fi
\expandafter\ifx\csname url\endcsname\relax
  \def\url#1{\texttt{#1}}\fi
\expandafter\ifx\csname urlprefix\endcsname\relax\def\urlprefix{URL }\fi
\providecommand{\bibinfo}[2]{#2}
\providecommand{\eprint}[2][]{\url{#2}}

\bibitem[{\citenamefont{Press et~al.}(2007)\citenamefont{Press, G{\"o}tzinger,
  Reitzenstein, Hofmann, L{\"o}ffler, Kamp, Forchel, and Yamamoto}}]{press:07}
\bibinfo{author}{\bibfnamefont{D.}~\bibnamefont{Press}},
  \bibinfo{author}{\bibfnamefont{S.}~\bibnamefont{G{\"o}tzinger}},
  \bibinfo{author}{\bibfnamefont{S.}~\bibnamefont{Reitzenstein}},
  \bibinfo{author}{\bibfnamefont{C.}~\bibnamefont{Hofmann}},
  \bibinfo{author}{\bibfnamefont{A.}~\bibnamefont{L{\"o}ffler}},
  \bibinfo{author}{\bibfnamefont{M.}~\bibnamefont{Kamp}},
  \bibinfo{author}{\bibfnamefont{A.}~\bibnamefont{Forchel}}, \bibnamefont{and}
  \bibinfo{author}{\bibfnamefont{Y.}~\bibnamefont{Yamamoto}},
  \bibinfo{journal}{Phys.\ Rev.\ Lett.} \textbf{\bibinfo{volume}{98}},
  \bibinfo{pages}{117402} (\bibinfo{year}{2007}).

\bibitem[{\citenamefont{Dousse et~al.}(2010)\citenamefont{Dousse,
  Suffczy{\'{n}}ski, Beveratos, Krebs, Lema{\^{i}}tre, Sagnes, Bloch, Voisin,
  and Senellart}}]{dousse:10}
\bibinfo{author}{\bibfnamefont{A.}~\bibnamefont{Dousse}},
  \bibinfo{author}{\bibfnamefont{J.}~\bibnamefont{Suffczy{\'{n}}ski}},
  \bibinfo{author}{\bibfnamefont{A.}~\bibnamefont{Beveratos}},
  \bibinfo{author}{\bibfnamefont{O.}~\bibnamefont{Krebs}},
  \bibinfo{author}{\bibfnamefont{A.}~\bibnamefont{Lema{\^{i}}tre}},
  \bibinfo{author}{\bibfnamefont{I.}~\bibnamefont{Sagnes}},
  \bibinfo{author}{\bibfnamefont{J.}~\bibnamefont{Bloch}},
  \bibinfo{author}{\bibfnamefont{P.}~\bibnamefont{Voisin}}, \bibnamefont{and}
  \bibinfo{author}{\bibfnamefont{P.}~\bibnamefont{Senellart}},
  \bibinfo{journal}{Nature} \textbf{\bibinfo{volume}{466}},
  \bibinfo{pages}{217} (\bibinfo{year}{2010}).

\bibitem[{\citenamefont{Bouwmeester et~al.}(2000)\citenamefont{Bouwmeester,
  Ekert, and Zeilinger}}]{bouwmeester:00}
\bibinfo{author}{\bibfnamefont{D.}~\bibnamefont{Bouwmeester}},
  \bibinfo{author}{\bibfnamefont{A.~K.} \bibnamefont{Ekert}}, \bibnamefont{and}
  \bibinfo{author}{\bibfnamefont{A.}~\bibnamefont{Zeilinger}},
  \emph{\bibinfo{title}{The Physics of Quantum Information}}
  (\bibinfo{publisher}{Springer}, \bibinfo{address}{Berlin},
  \bibinfo{year}{2000}).

\bibitem[{\citenamefont{Haroche and Raimond}(2006)}]{haroche:06}
\bibinfo{author}{\bibfnamefont{S.}~\bibnamefont{Haroche}} \bibnamefont{and}
  \bibinfo{author}{\bibfnamefont{J.}~\bibnamefont{Raimond}},
  \emph{\bibinfo{title}{Exploring the quantum}} (\bibinfo{publisher}{Oxford
  University Press}, \bibinfo{address}{Oxford}, \bibinfo{year}{2006}),
  \bibinfo{edition}{1st} ed.

\bibitem[{\citenamefont{Boeckler et~al.}(2008)\citenamefont{Boeckler,
  Reitzenstein, Kistner, Debusmann, Loeffler, Kida, Höfling, Forchel,
  Grenouillet, Claudon et~al.}}]{boeckler:08}
\bibinfo{author}{\bibfnamefont{C.}~\bibnamefont{Boeckler}},
  \bibinfo{author}{\bibfnamefont{S.}~\bibnamefont{Reitzenstein}},
  \bibinfo{author}{\bibfnamefont{C.}~\bibnamefont{Kistner}},
  \bibinfo{author}{\bibfnamefont{R.}~\bibnamefont{Debusmann}},
  \bibinfo{author}{\bibfnamefont{A.}~\bibnamefont{Loeffler}},
  \bibinfo{author}{\bibfnamefont{T.}~\bibnamefont{Kida}},
  \bibinfo{author}{\bibfnamefont{S.}~\bibnamefont{Höfling}},
  \bibinfo{author}{\bibfnamefont{A.}~\bibnamefont{Forchel}},
  \bibinfo{author}{\bibfnamefont{L.}~\bibnamefont{Grenouillet}},
  \bibinfo{author}{\bibfnamefont{J.}~\bibnamefont{Claudon}},
  \bibnamefont{et~al.}, \bibinfo{journal}{Appl.\ Phys.\ Lett.}
  \textbf{\bibinfo{volume}{92}}, \bibinfo{pages}{091107}
  (\bibinfo{year}{2008}).

\bibitem[{\citenamefont{Ates et~al.}(2009)\citenamefont{Ates, Ulrich, Ulhaq,
  Reitzenstein, L{\"o}ffler, H{\"o}fling, Forchel, and Michler}}]{ates:09}
\bibinfo{author}{\bibfnamefont{S.}~\bibnamefont{Ates}},
  \bibinfo{author}{\bibfnamefont{S.~M.} \bibnamefont{Ulrich}},
  \bibinfo{author}{\bibfnamefont{A.}~\bibnamefont{Ulhaq}},
  \bibinfo{author}{\bibfnamefont{S.}~\bibnamefont{Reitzenstein}},
  \bibinfo{author}{\bibfnamefont{A.}~\bibnamefont{L{\"o}ffler}},
  \bibinfo{author}{\bibfnamefont{S.}~\bibnamefont{H{\"o}fling}},
  \bibinfo{author}{\bibfnamefont{A.}~\bibnamefont{Forchel}}, \bibnamefont{and}
  \bibinfo{author}{\bibfnamefont{P.}~\bibnamefont{Michler}},
  \bibinfo{journal}{Nature Photonics} \textbf{\bibinfo{volume}{3}},
  \bibinfo{pages}{724} (\bibinfo{year}{2009}).

\bibitem[{\citenamefont{Kaer et~al.}(2010)\citenamefont{Kaer, Nielsen, Lodahl,
  Jauho, and M{\o}rk}}]{kaer:10}
\bibinfo{author}{\bibfnamefont{P.}~\bibnamefont{Kaer}},
  \bibinfo{author}{\bibfnamefont{T.~R.} \bibnamefont{Nielsen}},
  \bibinfo{author}{\bibfnamefont{P.}~\bibnamefont{Lodahl}},
  \bibinfo{author}{\bibfnamefont{A.~P.} \bibnamefont{Jauho}}, \bibnamefont{and}
  \bibinfo{author}{\bibfnamefont{J.}~\bibnamefont{M{\o}rk}},
  \bibinfo{journal}{Phys.\ Rev.\ Lett.} \textbf{\bibinfo{volume}{104}},
  \bibinfo{pages}{157401} (\bibinfo{year}{2010}).

\bibitem[{\citenamefont{Ulrich et~al.}(2011)\citenamefont{Ulrich, Ates,
  L{\"o}ffler, Forchel, and Michler}}]{ulrich:11}
\bibinfo{author}{\bibfnamefont{S.~M.} \bibnamefont{Ulrich}},
  \bibinfo{author}{\bibfnamefont{S.}~\bibnamefont{Ates},
  \bibfnamefont{S.~Reitzenstein}},
  \bibinfo{author}{\bibfnamefont{A.}~\bibnamefont{L{\"o}ffler}},
  \bibinfo{author}{\bibfnamefont{A.}~\bibnamefont{Forchel}}, \bibnamefont{and}
  \bibinfo{author}{\bibfnamefont{P.}~\bibnamefont{Michler}},
  \bibinfo{journal}{Phys.\ Rev.\ Lett.} \textbf{\bibinfo{volume}{106}},
  \bibinfo{pages}{247402} (\bibinfo{year}{2011}).

\bibitem[{\citenamefont{Roy and Hughes}(2011)}]{roy:11}
\bibinfo{author}{\bibfnamefont{C.}~\bibnamefont{Roy}} \bibnamefont{and}
  \bibinfo{author}{\bibfnamefont{S.}~\bibnamefont{Hughes}},
  \bibinfo{journal}{Phys.\ Rev.\ Lett.} \textbf{\bibinfo{volume}{106}},
  \bibinfo{pages}{247403} (\bibinfo{year}{2011}).

\bibitem[{\citenamefont{Kaer et~al.}(2013)\citenamefont{Kaer, Lodahl, Jauho,
  and M{\o}rk}}]{kaer:13a}
\bibinfo{author}{\bibfnamefont{P.}~\bibnamefont{Kaer}},
  \bibinfo{author}{\bibfnamefont{P.}~\bibnamefont{Lodahl}},
  \bibinfo{author}{\bibfnamefont{A.-P.} \bibnamefont{Jauho}}, \bibnamefont{and}
  \bibinfo{author}{\bibfnamefont{J.}~\bibnamefont{M{\o}rk}},
  \bibinfo{journal}{Phys.\ Rev.\ {\rm B}} \textbf{\bibinfo{volume}{87}},
  \bibinfo{pages}{081308} (\bibinfo{year}{2013}).

\bibitem[{\citenamefont{Harsij et~al.}(2012)\citenamefont{Harsij,
  Bagheri~Harouni, Roknizadeh, and Naderi}}]{harsij:12}
\bibinfo{author}{\bibfnamefont{Z.}~\bibnamefont{Harsij}},
  \bibinfo{author}{\bibfnamefont{M.}~\bibnamefont{Bagheri~Harouni}},
  \bibinfo{author}{\bibfnamefont{R.}~\bibnamefont{Roknizadeh}},
  \bibnamefont{and} \bibinfo{author}{\bibfnamefont{M.~H.}
  \bibnamefont{Naderi}}, \bibinfo{journal}{Phys.\ Rev.\ A}
  \textbf{\bibinfo{volume}{86}}, \bibinfo{pages}{063803}
  (\bibinfo{year}{2012}).

\bibitem[{\citenamefont{Gl{\"a}ssl et~al.}(2012)\citenamefont{Gl{\"a}ssl,
  S{\"o}rgel, Vagov, Croitoru, Kuhn, and Axt}}]{glaessl:12b}
\bibinfo{author}{\bibfnamefont{M.}~\bibnamefont{Gl{\"a}ssl}},
  \bibinfo{author}{\bibfnamefont{L.}~\bibnamefont{S{\"o}rgel}},
  \bibinfo{author}{\bibfnamefont{A.}~\bibnamefont{Vagov}},
  \bibinfo{author}{\bibfnamefont{M.~D.} \bibnamefont{Croitoru}},
  \bibinfo{author}{\bibfnamefont{T.}~\bibnamefont{Kuhn}}, \bibnamefont{and}
  \bibinfo{author}{\bibfnamefont{V.~M.} \bibnamefont{Axt}},
  \bibinfo{journal}{Phys.\ Rev.\ {\rm B}} \textbf{\bibinfo{volume}{86}},
  \bibinfo{pages}{035319} (\bibinfo{year}{2012}).

\bibitem[{\citenamefont{McCutcheon and Nazir}(2013)}]{nazir:13}
\bibinfo{author}{\bibfnamefont{D.~P.~S.} \bibnamefont{McCutcheon}}
  \bibnamefont{and} \bibinfo{author}{\bibfnamefont{A.}~\bibnamefont{Nazir}},
  \bibinfo{journal}{Phys.\ Rev.\ Lett.} \textbf{\bibinfo{volume}{110}},
  \bibinfo{pages}{217401} (\bibinfo{year}{2013}).

\bibitem[{\citenamefont{Breuer and Petruccione}(2002)}]{breuer:02}
\bibinfo{author}{\bibfnamefont{H.~P.} \bibnamefont{Breuer}} \bibnamefont{and}
  \bibinfo{author}{\bibfnamefont{F.}~\bibnamefont{Petruccione}},
  \emph{\bibinfo{title}{The Theory of Open Quantum Systems}}
  (\bibinfo{publisher}{Oxford University Press}, \bibinfo{address}{Oxford},
  \bibinfo{year}{2002}), \bibinfo{edition}{1st} ed.

\bibitem[{\citenamefont{Mahan}(1990)}]{mahan:90}
\bibinfo{author}{\bibfnamefont{G.~D.} \bibnamefont{Mahan}},
  \emph{\bibinfo{title}{Many-Particle Physics}} (\bibinfo{publisher}{Plenum
  Press}, \bibinfo{address}{New York}, \bibinfo{year}{1990}),
  \bibinfo{edition}{2nd} ed.

\bibitem[{\citenamefont{Mukamel}(1995)}]{mukamel:95}
\bibinfo{author}{\bibfnamefont{S.}~\bibnamefont{Mukamel}},
  \emph{\bibinfo{title}{Principles of Nonlinear Optical Spectroscopy}}
  (\bibinfo{publisher}{Oxford University Press}, \bibinfo{address}{New York,
  Oxford}, \bibinfo{year}{1995}).

\bibitem[{\citenamefont{Krummheuer et~al.}(2002)\citenamefont{Krummheuer, Axt,
  and Kuhn}}]{krummheuer:02}
\bibinfo{author}{\bibfnamefont{B.}~\bibnamefont{Krummheuer}},
  \bibinfo{author}{\bibfnamefont{V.~M.} \bibnamefont{Axt}}, \bibnamefont{and}
  \bibinfo{author}{\bibfnamefont{T.}~\bibnamefont{Kuhn}},
  \bibinfo{journal}{Phys.\ Rev.\ {\rm B}} \textbf{\bibinfo{volume}{65}},
  \bibinfo{pages}{195313} (\bibinfo{year}{2002}).

\bibitem[{\citenamefont{Andreani et~al.}(1999)\citenamefont{Andreani,
  Panzarini, and Gerard}}]{andreani:99}
\bibinfo{author}{\bibfnamefont{L.C.}~\bibnamefont{Andreani}},
  \bibinfo{author}{\bibfnamefont{G.}~\bibnamefont{Panzarini}},
  \bibnamefont{and} \bibinfo{author}{\bibfnamefont{J.~M.}
  \bibnamefont{Gerard}}, \bibinfo{journal}{Phys.\ Rev.\ {\rm B}}
  \textbf{\bibinfo{volume}{60}}, \bibinfo{pages}{13276} (\bibinfo{year}{1999}).

\bibitem[{\citenamefont{Rudin and Reinecke}(1999)}]{rudin:99}
\bibinfo{author}{\bibfnamefont{S.}~\bibnamefont{Rudin}} \bibnamefont{and}
  \bibinfo{author}{\bibfnamefont{T.~L.} \bibnamefont{Reinecke}},
  \bibinfo{journal}{Phys.\ Rev.\ {\rm B}} \textbf{\bibinfo{volume}{59}},
  \bibinfo{pages}{10227} (\bibinfo{year}{1999}).

\bibitem[{\citenamefont{Gl{\"a}ssl et~al.}(2011)\citenamefont{Gl{\"a}ssl,
  Vagov, L{\"u}ker, Reiter, Croitoru, Machnikowski, Axt, and
  Kuhn}}]{glaessl:11b}
\bibinfo{author}{\bibfnamefont{M.}~\bibnamefont{Gl{\"a}ssl}},
  \bibinfo{author}{\bibfnamefont{A.}~\bibnamefont{Vagov}},
  \bibinfo{author}{\bibfnamefont{S.}~\bibnamefont{L{\"u}ker}},
  \bibinfo{author}{\bibfnamefont{D.~E.} \bibnamefont{Reiter}},
  \bibinfo{author}{\bibfnamefont{M.~D.} \bibnamefont{Croitoru}},
  \bibinfo{author}{\bibfnamefont{P.}~\bibnamefont{Machnikowski}},
  \bibinfo{author}{\bibfnamefont{V.~M.} \bibnamefont{Axt}}, \bibnamefont{and}
  \bibinfo{author}{\bibfnamefont{T.}~\bibnamefont{Kuhn}},
  \bibinfo{journal}{Phys.\ Rev.\ {\rm B}} \textbf{\bibinfo{volume}{84}},
  \bibinfo{pages}{195311} (\bibinfo{year}{2011}).

\bibitem[{\citenamefont{Vagov et~al.}(2011)\citenamefont{Vagov, Croitoru,
  Gl{\"a}ssl, Axt, and Kuhn}}]{vagov:11}
\bibinfo{author}{\bibfnamefont{A.}~\bibnamefont{Vagov}},
  \bibinfo{author}{\bibfnamefont{M.~D.} \bibnamefont{Croitoru}},
  \bibinfo{author}{\bibfnamefont{M.}~\bibnamefont{Gl{\"a}ssl}},
  \bibinfo{author}{\bibfnamefont{V.~M.} \bibnamefont{Axt}}, \bibnamefont{and}
  \bibinfo{author}{\bibfnamefont{T.}~\bibnamefont{Kuhn}},
  \bibinfo{journal}{Phys.\ Rev.\ {\rm B}} \textbf{\bibinfo{volume}{83}},
  \bibinfo{pages}{094303} (\bibinfo{year}{2011}).

\bibitem[{\citenamefont{Reithmaier et~al.}(2004)\citenamefont{Reithmaier, Sek,
  L{\"o}ffler, Hofmann, Kuhn, Reitzenstein, Keldysh, Kulakovskii, Reinecke, and
  Forchel}}]{reithmaier:04}
\bibinfo{author}{\bibfnamefont{J.~P.} \bibnamefont{Reithmaier}},
  \bibinfo{author}{\bibfnamefont{G.}~\bibnamefont{Sek}},
  \bibinfo{author}{\bibfnamefont{A.}~\bibnamefont{L{\"o}ffler}},
  \bibinfo{author}{\bibfnamefont{C.}~\bibnamefont{Hofmann}},
  \bibinfo{author}{\bibfnamefont{S.}~\bibnamefont{Kuhn}},
  \bibinfo{author}{\bibfnamefont{S.}~\bibnamefont{Reitzenstein}},
  \bibinfo{author}{\bibfnamefont{L.~V.} \bibnamefont{Keldysh}},
  \bibinfo{author}{\bibfnamefont{V.~D.} \bibnamefont{Kulakovskii}},
  \bibinfo{author}{\bibfnamefont{T.~L.} \bibnamefont{Reinecke}},
  \bibnamefont{and} \bibinfo{author}{\bibfnamefont{A.}~\bibnamefont{Forchel}},
  \bibinfo{journal}{Nature} \textbf{\bibinfo{volume}{432}},
  \bibinfo{pages}{197} (\bibinfo{year}{2004}).

\bibitem[{\citenamefont{Hennessy et~al.}(2007)\citenamefont{Hennessy, Badolato,
  Winger, Gerace, Atat{\"u}re, Gulde, F{\"a}lt, Hu, and
  Imamo{\v{g}}lu}}]{hennessy:07}
\bibinfo{author}{\bibfnamefont{K.}~\bibnamefont{Hennessy}},
  \bibinfo{author}{\bibfnamefont{A.}~\bibnamefont{Badolato}},
  \bibinfo{author}{\bibfnamefont{M.}~\bibnamefont{Winger}},
  \bibinfo{author}{\bibfnamefont{D.}~\bibnamefont{Gerace}},
  \bibinfo{author}{\bibfnamefont{M.}~\bibnamefont{Atat{\"u}re}},
  \bibinfo{author}{\bibfnamefont{S.}~\bibnamefont{Gulde}},
  \bibinfo{author}{\bibfnamefont{S.}~\bibnamefont{F{\"a}lt}},
  \bibinfo{author}{\bibfnamefont{E.~L.} \bibnamefont{Hu}}, \bibnamefont{and}
  \bibinfo{author}{\bibfnamefont{A.}~\bibnamefont{Imamo{\v{g}}lu}},
  \bibinfo{journal}{Nature} \textbf{\bibinfo{volume}{445}},
  \bibinfo{pages}{896} (\bibinfo{year}{2007}).

\bibitem[{\citenamefont{Machnikowski and Jacak}(2004)}]{machnikowski:04}
\bibinfo{author}{\bibfnamefont{P.}~\bibnamefont{Machnikowski}}
  \bibnamefont{and} \bibinfo{author}{\bibfnamefont{L.}~\bibnamefont{Jacak}},
  \bibinfo{journal}{Phys.\ Rev.\ {\rm B}} \textbf{\bibinfo{volume}{69}},
  \bibinfo{pages}{193302} (\bibinfo{year}{2004}).

\bibitem[{\citenamefont{Vagov et~al.}(2007)\citenamefont{Vagov, Croitoru, Axt,
  Kuhn, and Peeters}}]{vagov:07}
\bibinfo{author}{\bibfnamefont{A.}~\bibnamefont{Vagov}},
  \bibinfo{author}{\bibfnamefont{M.~D.} \bibnamefont{Croitoru}},
  \bibinfo{author}{\bibfnamefont{V.~M.} \bibnamefont{Axt}},
  \bibinfo{author}{\bibfnamefont{T.}~\bibnamefont{Kuhn}}, \bibnamefont{and}
  \bibinfo{author}{\bibfnamefont{F.~M.} \bibnamefont{Peeters}},
  \bibinfo{journal}{Phys.\ Rev.\ Lett.} \textbf{\bibinfo{volume}{98}},
  \bibinfo{pages}{227403} (\bibinfo{year}{2007}).

\end{thebibliography}

\end{document}